# Pattern Formation Mechanism of Directionally-Solidified MoSi$_2$/Mo$_5$Si$_3$ Eutectic by Phase-Field Simulation

Chuanqi Zhu[a†], Yuichiro Koizumi[b*], Akihiko Chiba[c],
Koretaka Yuge[d], Kyosuke Kishida[d], Haruyuki Inui[d]

[a]*Graduate School of Engineering, Tohoku University*
[b]*Division of Materials and Manufacturing Science, Osaka University*
[c]*Institute for Materials Research, Tohoku University*
[d]*Center for Elements Strategy Initiative for Structure Materials (ESISM), Kyoto University*

*Corresponding author.
Email-address: ykoizumi@mat.eng.osaka-u.ac.jp (Y.Koizumi)
† Present address: Division of Materials and Manufacturing Science, Osaka University

**Abstract**

A phase-field study has been conducted to obtain an understanding of the formation mechanism of the script lamellar pattern of MoSi$_2$/Mo$_5$Si$_3$ eutectic composite, which is a promising candidate for high-temperature structural application. The spacing of the lamellar pattern in the simulation results shows good agreement with that of experimental observations and analytical solutions under three growth rates: 10 mm·h$^{-1}$, 50 mm·h$^{-1}$, and 100 mm·h$^{-1}$. The discontinuity of Mo$_5$Si$_3$ rods, in contrast to the regular eutectic with a continuous pattern, is claimed to be caused by the instability of the solid-liquid interface. In this study, the implementation of Mo$_5$Si$_3$ nucleation over the solid-liquid interface has been proposed and successfully reproduced the characteristic of discontinuity. A highly random and intersected lamellar pattern similar to that observed in the ternary MoSi$_2$/Mo$_5$Si$_3$ eutectic alloyed with 0.1at% Co has been obtained in simulation owing to the increase in the frequency of nucleation. In addition, it has been demonstrated that the inclination of the Mo$_5$Si$_3$ rod can be reproduced by taking into account the strong relaxation of lattice strain energy, which is generally considered to be negligible in eutectic reaction, as the result of the formation of ledge-terrace structure.

## 1. Introduction

MoSi$_2$-based alloys have been considered as promising candidates for high-temperature structural application owing to its high melting point, good oxidation resistance, relatively low density [1-4]. Directionally-solidified MoSi$_2$/Mo$_5$Si$_3$ spontaneous composite is of particular interest because of its better creep property than other MoSi$_2$-based composites [5][6]. However, its fracture toughness at room temperature is still insufficient for practical application. As this shortage is believed to be overcome by microstructure refinement and morphology control [6][7], it is quite important to elucidate the formation mechanism of the MoSi$_2$/Mo$_5$Si$_3$ microstructure, which has an exotic eutectic pattern, the so-called script lamellar pattern (Fig. 1a). In this study, the phase-field method has been firstly used to reveal the underlying physics during the solidification process of MoSi$_2$/Mo$_5$Si$_3$ eutectic.

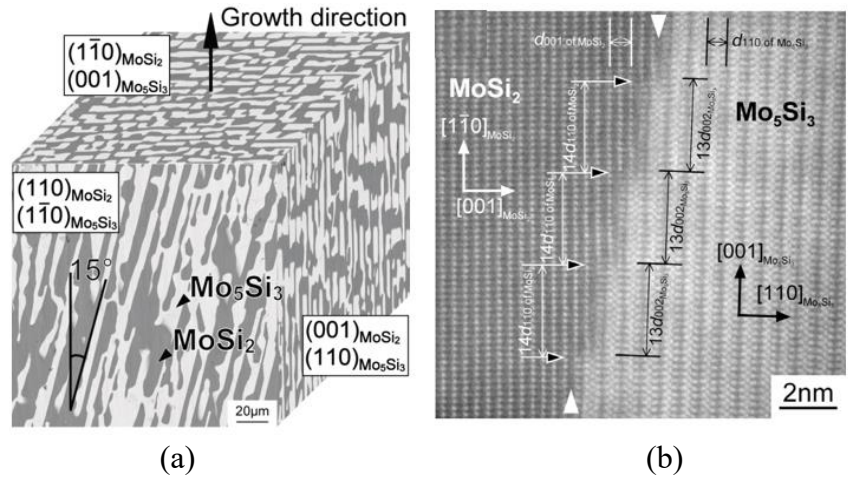

(a) (b)

Fig.1 (a) Three-dimensional view of script lamellar pattern of MoSi$_2$/Mo$_5$Si$_3$ eutectic, reconstructed from images. The dark phase is MoSi$_2$ matrix and the grey are Mo$_5$Si$_3$ rods. (b) Inclined interface with ledge-terrace structure resolved at atomic scale observed from (110) of MoSi$_2$ and (1-10) of Mo$_5$Si$_3$ (after Fujiwara et al. [7]).

Phase-field method has emerged as the most powerful tool for simulation of pattern formation in solidification [8][9]. In the phase-field method, order parameters are defined to represent the microstructure and act as variables for the total free energy of the system. The values of order parameters vary continuously in the interface region, forming a smooth and diffuse profile, while it keeps constant in the bulk regions (i.e., within grains, domains, phases, and so on). This implicit representation of microstructure by the diffuse interface makes it possible to model complex pattern formation in solidification, such as dendrite structure [10]. Phase-field equation governs the microstructure evolution and is derived by the variational differentiation of the total free energy, based on the principle of energy minimization. The total free energy can be described as the sum of chemical energy, interfacial energy, elastic energy. By using physical properties evaluated by other approaches, such as the CALPHAD method, First Principle calculation or experimental measurement, it is possible to perform quantitative or quasi-quantitative phase-field simulation [11]. The insights gained from the simulation will provide a deep understanding of material behavior, which is crucial for the optimization of processing parameters and material performance [12]. In order to conduct a phase-field simulation of the MoSi$_2$/Mo$_5$Si$_3$ eutectic, it is necessary to start with a discussion on previous theoretical and experimental studies of eutectic solidification.

The process of eutectic solidification, in which the liquid phase transforms into at least two solid phases simultaneously, has been extensively studied. The cornerstone theory for understanding eutectic growth was established by Jackson and Hunt [13]. The theory pointed out the interplay between the diffusion effect and interface tension. It is stated that for a given cooling rate, a steady-state spacing of eutectic lamellae shall be realized by the balance of solute undercooling and curvature undercooling, reaching a minimum value of total undercooling. Thus, this theory provides a guideline for predicting the lamellar spacing and regulating the fineness of lamellar structure by adjusting the cooling rate. Moreover, the morphologies of eutectics are mainly determined by the crystallographic properties of the solid phases. For example, the coupling of a facet and a non-facet phase gives rise to the formation of irregular eutectic structure, having random, discontinuous phase distribution [14]. Furthermore, the anisotropy in the interfacial energy of the solid-solid interface can lead to a large variety of unique and intriguing eutectic morphologies [15].

A noticeable characteristic of the $MoSi_2/Mo_5Si_3$ eutectic microstructure is the discontinuity of the $Mo_5Si_3$ rods embedded in the $MoSi_2$ matrix, which is similar to that for the typical irregular eutectic pattern of Al-Si (Fig.1b in [14]) alloy. The mechanism of discontinuous pattern in irregular eutectic was investigated by experimental observation. It has been found that the solid-liquid interface has an uneven and unstable morphology (Fig.1d in [14]) and is in a non-isothermal state. This instability of interface is mainly because the facet phase cannot follow the growth of the non-facet matrix; thus, they cannot form a stable coupled growth. In contrast to the non-faceted interface of Al crystal with metallic bond, the Si crystal with covalent bond tends to form faceted interface by growing along a specific crystallographic direction [16], exhibiting strong anisotropy of interface mobility. The preferential growth direction of each Si grain relative to the temperature gradient is random and determined by its initial orientation at the stage of nucleation. Therefore, the Si grains with random growth directions keep being overgrown by the matrix Al-phase and emerging as nuclei, causing an instability of the solid-liquid interface. Analogously, there is no doubt that the solid-liquid interface of $MoSi_2/Mo_5Si_3$ eutectic with the discontinuous pattern is also unstable during the directional solidification, principally owing to the directional bonding of the intermetallic compound.

Another irregular characteristic of the $MoSi_2/Mo_5Si_3$ script lamellar pattern is the inclined solid-solid interface relative to the temperature gradient. The inclination phenomenon of the eutectic pattern has been previously reported in the literature [17][18]. Generally, inclined patterns can be induced by the effect of either diffusion or interface energy anisotropy. The inclination induced by diffusion occurs in thin-film samples or two-dimension simulations [18]. When the solute diffusion is limited compared to the growth rate, eutectic lamellae grow in an angle with respect to the direction of the temperature gradient for the sake of facilitating solute exchange between the two coupled phases. The inclination angle is determined by the balance of growth rate and diffusion rate. In the case of another effect, i.e., anisotropy in interface anisotropy, the inclined interface is fixed to a direction determined by the crystallographic relationship between the two coupled phases. For $MoSi_2/Mo_5Si_3$ eutectic, a ledge-terrace structure has been observed at the inclined interface by high-resolution TEM (Fig.1b), which is a sign of lattice accommodation for reduction of elastic energy. Therefore, the inclination of $MoSi_2/Mo_5Si_3$ interface is much more likely to be caused by the anisotropy of solid-solid interface energy. However, the influence of diffusion cannot completely be excluded, and a rigorous discussion is needed to figure out the main factor for the inclined morphology.

Accordingly, the possible formation mechanisms of the irregular and complex $MoSi_2/Mo_5Si_3$ script lamellar pattern are: (I) formation of discontinuous rods led by the instability of the solid-liquid interface; (II) interface inclination caused by anisotropy of the solid-solid interface energy; (III) interface inclination induced by diffusion in the liquid. It is important to examine the contribution of these factors for the control of microstructure and the improvement of the mechanical properties. In the present study, the roles of these factors in the formation of the script lamellar structure are examined in the following steps: Firstly, a regular eutectic solidification model is constructed. The front undercooling at steady state extracted from the two-dimension phase-field simulation results needs to be compared with that in the analytical results, which has direct correlations to material properties. The PF model should also be able to reproduce the spacing transition phenomena observed in the experiment and predicted by theory [13]. Secondly, the two-dimension model is extended to a three-dimension one in order to output microstructure images comparable to experimental results in [6]. Then, a nucleation functionality is incorporated into this eutectic model. It is assumed that nucleation is a representation of solid-liquid interface instability, which is responsible for the pattern discontinuity. Finally, in order to find out the primary factor that determines the inclined morphology of script lamellar pattern, another phase-field model with anisotropic interface energy is developed. In combination with the experimental evidence, attempts are made to obtain an anisotropic description of the solid-solid interface energy. Overall, the simulation results of these steps are expected to provide a new perspective to understand the eutectic solidification process of the $MoSi_2/Mo_5Si_3$ intermetallic composite.

## 2. Methods

### 2.1 Analytical model for undercooling of a regular, lamellar eutectic front

The following analytical method for coupled growth of eutectic lamellae was proposed by Jackson and Hunt [13] and adapted by Dantzig et al. [20]. Starting with the Gibbs-Thomson equation, the interface undercooling can be given as,

$$\Delta T = \Delta T_{curvature} + \Delta T_{solute} \tag{1}$$

which sums up the curvature undercooling and solute undercooling without consideration of the kinetic undercooling on the assumption of low solidification rate. By solving the diffusion equation and averaging curvature of the eutectic front, Jackson and Hunt derived the undercooling with relation to the growth velocity $v$ and the lamellar spacing $\lambda$ for a binary eutectic system as,

$$\Delta T = A_1 \frac{v}{D_l} \lambda + A_2 \frac{1}{\lambda} \tag{2}$$

in which $A_1$ and $A_2$ are coefficients with relation to thermodynamic and interface properties, such as phase equilibrium, fusion entropy, and interface energy. The detailed properties of $MoSi_2/Mo_5Si_3$ eutectic system are listed in Table 1. They can be expressed as,

$$A_1 = \frac{\Delta C_0}{g_\alpha g_\beta} \frac{|m_{l\alpha}||m_{l\beta}|}{|m_{l\alpha}|+|m_{l\beta}|} \sum_{n=1}^{\infty} \frac{\sin^2(n\pi g_\alpha)}{(n\pi)^3} \tag{3}$$

$$A_2 = \frac{|m_{l\alpha}||m_{l\beta}|}{|m_{l\alpha}|+|m_{l\beta}|} \left(\frac{2\gamma_{\alpha l} \cos\theta_\alpha}{|m_{l\alpha}|g_\alpha \Delta S_{\alpha l}} + \frac{2\gamma_{\beta l} \cos\theta_\beta}{|m_{l\beta}|g_\beta \Delta S_{\beta l}}\right) \tag{4}$$

For three different cooling rates (10 mm·h$^{-1}$, 50 mm·h$^{-1}$, and 100 mm·h$^{-1}$), the interface undercooling is plotted as a function of lamellar spacing. These plots act as the baselines for examining the validity of the phase-field model specified as below.

### 2.2 Phase-Field Model

The commercially available software MICRESS (version 6.300) [21], based on the multi-phase-field framework [22], has been used for conducting a phase-field simulation of eutectic solidification. Three non-conservative order parameters ($\phi_0$, $\phi_1$, $\phi_2$) were assigned for liquid, $MoSi_2$, and $Mo_5Si_3$ phases fields, and one conservative order parameter $c$ was assigned for Si concentration in the eutectic system. The phase-field equation governing the interface kinetics is expressed as,

$$\frac{\partial \phi_\alpha}{\partial t} = \sum_{\beta=1}^{v} M_{\alpha\beta} \left[ \sigma_{\alpha\beta} (\nabla^2 \phi_\alpha - \nabla^2 \phi_\beta) + \frac{\pi^2}{2\eta^2} (\phi_\alpha - \phi_\beta) + \frac{\pi}{\eta} \sqrt{\phi_\alpha \phi_\beta} \Delta G_{\alpha\beta} \right] \quad (5)$$

in which the evolution of phase α is the result of interaction with all surrounding phases. $M_{\alpha\beta}$, $\sigma_{\alpha\beta}$ and $\Delta G_{\alpha\beta}$ are interface mobility, interface energy and driving force between two phases of α and β. The artificially large interface width $\eta$ is used in simulaitn for computation efficiency and set to be three times grid size for proper interface resolution. The driving force can be expressed as,

$$\Delta G_{\alpha\beta} = \Delta S_{\alpha\beta} \Delta T(c) \quad (6)$$

$\Delta S_{\alpha\beta}$ is the fusion entropy between α and β phase. $\Delta T(c)$ is the difference between the actual temperature in the system and the liquidus temperature for the concentration $c$. The diffusion equation controlling the solute redistribution in solid and liquid phases can be written as

$$\frac{\partial c}{\partial t} = \nabla (\sum_{\alpha=1}^{v} \phi_\alpha D_\alpha \nabla c_\alpha) \quad (7)$$

in which $c_\alpha$ and $D_\alpha$ are concentration field and the diffusion coefficient in the α phase, respectively. The physical properties of $MoSi_2$/$Mo_5Si_3$ eutectic system in this phase-field model are the same as those used in the analytical method (see Table 1).

Table 1 Physical parameters used in analytical and phase field model

| Item | Symbol | Value |
|---|---|---|
| Liquids Slope | $m_{l\alpha}$ | -3100 K |
| | $m_{l\beta}$ | 1800 K |
| Liquid concentration | $C_l$ | 54 at.% |
| Concentration Difference of solid phases | $\Delta C_0$ | 0.28 |
| Phase Fraction (α-$Mo_5Si_3$, β-$MoSi_2$) | $g_\alpha$ | 0.393 |
| | $g_\beta$ | 0.607 |
| Contact Angle | $\theta_\alpha$ | $3\pi/8$ |
| | $\theta_\beta$ | $3\pi/8$ |
| Fusion Entropy | $\Delta S$ | $5 \times 10^6$ J·K$^{-1}$·m$^{-3}$ |
| Diffusivity in Liquid | $D_l$ | $2.45 \times 10^{-8}$ m$^2$·s$^{-1}$ |
| Interface Energy | $\gamma_{\alpha l}, \gamma_{\beta l}$ | 1.0 J·m$^{-2}$ |

For 2D simulation, the mesh number of the box in the height direction is set as 200 with a grid size of 0.1μm. With a single pair of lamellae in the box, the lamellar spacing can be set by changing the mesh number of the box in the width direction. The temperature profile is demonstrated by the DP MICRESS software. The resultant interface undercooling is measured and compared with the analytical solutions. In the case of 3D simulation extended from the 2D model, the mesh number of the box is 50×50×100 with a grid size of 0.4μm. The simulated microstructure pattern is visualized by an open-source software ParaView (Kitware). For quantifying the eutectic structure, the lamellar spacing of both simulation and experiment images are analyzed by Fiji, an open-source image analysis package based on Image-J. The "particle analysis" is used to obtain the barycenter coordinate information of rods in the azimuthal (perpendicular to growth direction) cross section. An algorithm programmed in MatLab is used to determine the distance to the nearest neighbor for one rod by calculating the distances to all the other rods and choosing the minimum value. This process was repeated one by one for each rod. The lamellar spacing is obtained by averaging the values of the nearest neighbor distances.

In addition, another multi-phase-field model with a different formulation incorporated with anisotropic interface energy is employed to examine the factors for interface inclination. For the convenience of implementation, the model is derived in a different formulation and uses a set of non-dimensionalized parameters (Appendix Table). The detailed derivation of this model is presented in [18]. In brief, the phase-field equation is given as

$$\frac{\partial \phi_\alpha}{\partial t} = -\frac{2}{n}\sum_{j=1}^{n} M_{ij}\left[\sum_{k=1}^{n}\frac{1}{2}(\varepsilon_{ik}^2 - \varepsilon_{jk}^2)\nabla^2\phi_k + (W_{ik} - W_{jk})\phi_k - \frac{8}{\pi}\sqrt{\phi_i\phi_j}\Delta f_{ij}\right] \quad (8)$$

The parameters $M_{ij}$, $\varepsilon_{ij}$, $W_{ij}$, and $\Delta f_{ij}$ are matrix quantities representing phase interactions between neighbor phases at interfaces. $M_{ij}$ is the interface mobility, $\varepsilon_{ij}$ is the gradient energy coefficient, $W_{ij}$ is the magnitude of energy penalty for the interface, and $\Delta f_{ij}$ is the driving force calculated from the chemical free energy difference between the two phases. For describing anisotropic interface energy, the coefficient $\varepsilon_{ij}$ is set to be anisotropic and given as a function of interface normal angle $\theta$. Thus, in the two-dimension coordinate system, the Laplacian terms $\varepsilon_{ij}^2\nabla^2\phi_j$ in the Eq.8 are replaced by

$$\nabla\left[\varepsilon(\theta)_{ij}^2\nabla\phi_j\right] - \frac{\partial}{\partial x}\left[\varepsilon(\theta)_{ij}\frac{\partial\varepsilon(\theta)_{ij}}{\partial\theta_{ij}}\frac{\partial\phi_j}{\partial y}\right] + \frac{\partial}{\partial y}\left[\varepsilon(\theta)_{ij}\frac{\partial\varepsilon(\theta)_{ij}}{\partial\theta_{ij}}\frac{\partial\phi_j}{\partial x}\right] \quad (9)$$

The following two anisotropy functions of ε coefficient with different features are used in the present work,

$$\varepsilon(\theta) = 1 - \delta\cos[k(\theta + \theta_0)] \quad (10)$$

$$\varepsilon(\theta) = 1 - \delta\exp\left[-\frac{\sin^2(\theta + \theta_0)}{w^2}\right] \quad (11)$$

The parameters $\delta$, $k$, $\theta_0$, and $w$ in these functions are used to adjust the shape of their polar plots. The values of interface inclination resulting from introducing these different anisotropy functions are compared to discuss the anisotropy property of the interface.

## 3. Model Verification

*3.1 Analysis of interface undercooling*

The 2D simulation box is illustrated in Fig.2: a positive linear temperature gradient is distributed from the left to the right side; the domain is cooled by moving this gradient at a constant velocity. After reaching a steady state, a pair of lamellae, with a value of spacing equal to the box width, grows in a constant velocity equal to that of temperature gradient. Both of the PF model and the JH model use the same set of physical parameters listed in Table 1. The parameters related to phase diagram are extracted from thermodynamic data in [23]. Based on the empirical rule that the fusion entropy per mole should be larger than 2R (R is the gas constant) for facet crystal like Molybdenum, the fusion entropy is around the value of $1.7×10^6$ J·K$^{-1}$·m$^{-3}$. Some experimental measurement has estimated the interface energy of Molybdenum to be around 0.8J·m$^{-2}$[24]. Considering the highly ordered crystallographic structure of both MoSi$_2$ and Mo$_5$Si$_3$ phases, the fusion entropy and interface energy are preliminarily determined as $5×10^6$ J·K$^{-1}$·m$^{-3}$ and 1.0J·m$^{-2}$. By reference to the diffusivity of Silicon in aluminum which is in the order of -8[25], the diffusivity of Silicon in Molybdenum is determined as $2.45×10^{-8}$ m$^2$·s$^{-1}$ by manually changing the value of diffusivity in this order until the spacing in simulation results become nearly consistent with that of experiment results. However, it is not guaranteed that this set of parameters can be the actual representative of the MoSi$_2$/Mo$_5$Si$_3$ eutectic system. The determination of these parameters is still in discussion.

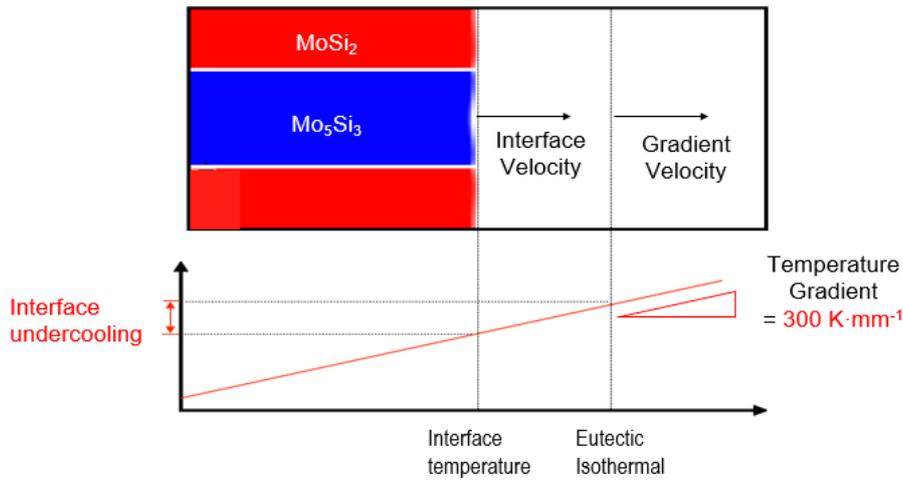

Fig. 2 Simulation box and temperature profile in two dimensions. The domain is cooled by moving the temperature gradient at a constant velocity. The interface undercooling reach a stable value when the interface velocity equals the gradient velocity at steady state.

In Fig.3, for three steady-state interface velocities (10 mm·h$^{-1}$, 50 mm·h$^{-1}$, 100 mm·h$^{-1}$), the interface undercooling results from PF simulation (circles interpolated by dash lines) are plotted together with the Jackson-Hunt (JH) analytical solutions (solid lines). The curves of PF results are found to have the same valley-like shape as the JH solutions. The values of extreme lamellar spacing (denoted by arrows) from PF results are close to those from JH results for every interface velocity. As summarized in Table 2, the deviations between them vary in the range of 10% to 17%. However, the extreme undercooling in PF results is obviously larger than that in JH results. This is because the kinetic undercooling is neglected in the derivation of Eq. 1 and Eq. 2. However, this difference does not matter for the relatively low cooling rate in our cases. The kinetic effect is less important than the interaction between diffusion and interface tension, which determines the length scale of the eutectic structure. By the undercooling comparison above, the phase-field model is able to be put into further development for simulating the spacing transition phenomenon of the lamellar structure.

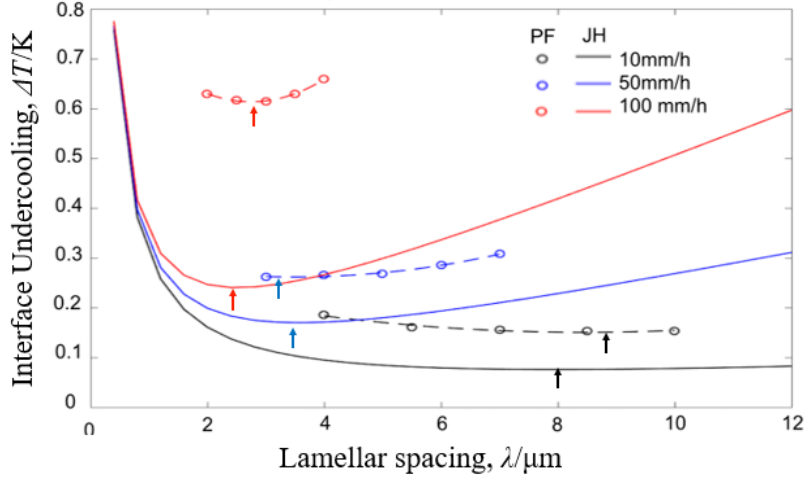

Fig.3 Comparison of interface undercooling between phase-field (dash lines) and Jackson-Hunt (solid lines) results at three interface velocities (10 mm/h, 50 mm/h, 100 mm/h).

Table 2 Comparison of extreme spacing results from Jackson-Hunt (JH) solution and two-dimension phase-field (PF) simulation.

|  | Interface Velocity, $v_{int}$ (mm·h$^{-1}$) | | |
| --- | --- | --- | --- |
|  | 10 | 50 | 100 |
|  | Extreme Spacing, μm | | |
| JH Solution | 8.00 | 3.60 | 2.40 |
| PF Simulation | 8.80 | 3.20 | 2.80 |
| Deviation | 10.0% | 11.1% | 16.7% |

*3.2 Reproduction of spacing transition*

Spacing transition phenomenon has been experimentally observed in transparent eutectic alloy [13] and also simulated by numerical methods [27]. The lamellar structure will be unstable when their spacing deviates far from the extreme condition for a given steady state interface velocity. It tends to have a more stable spacing through a transition process. In order to reproduce this process on the basis of the previously developed eutectic model with a single pair of lamellae, we extend it to the case of multiple pairs with various magnitudes of spacing.

In Fig.4, all the PF simulations were performed at the interface velocity of 100 mm·h$^{-1}$, corresponding to the JH solution denoted by the red line in Fig.3. As Fig.4b shows, when the initial spacing (2.5 μm) is close to the extreme spacing (2.4 μm) denoted in Fig.3, although there is a small fluctuation at the beginning of the growth, all the lamellae survive and grow in a steady state eventually. In Fig.4a, the initial spacing (0.625 μm) is much smaller than the extreme spacing. Right after the start, some lamellae are overgrown by adjacent ones. As a result, the lamellar spacing transforms to be near the extreme spacing. In Fig.4c, the initial lamellar spacing is very large, and the lamellae eventually grow in an inclined angle. This instability phenomenon attracts our particular attention. The inclination observed here is surely not caused by the interface energy anisotropy, which is not included in the model. The main reason is claimed to be the spatial restriction in the two-dimension box. When the 2D model in Fig.4c is extended to a 3D one, as shown in Fig.4d,e, the rod phase in the center splits into several branches, resulting in the reduction of the spacing and the stabilization of the lamellar structure.

By analyzing the solute distribution ahead of the eutectic front in Fig.4c, more insights into the space-restricted instability can be obtained. Basically, the solute-rich phase and solvent-rich phase of eutectic need solute and solvent atoms for their growth, respectively. The atoms are redistributed by diffusion ahead of the solid-liquid interface. Owing to the large initial lamellar spacing, the element ejected by one phase cannot easily transport to the adjacent counter phase. The element exchange between the coupled phases is limited, resulting in the massive accumulation or depletion of the solute concentration, as indicated by the concentration contrast ahead of the interface in Fig.4c. As a result, the interface forms a concave shape, which is the initial stage of splitting. However, due to the geometrical limitation of the two-dimension box, the splitting in the case of 3D simulation (Fig.4d,e) cannot develop. For the sake of alleviation of the concentration accumulation or depletion, the lamellae grow in an inclined angle to facilitate the solute redistribution. From the examination above, this model is expected to produce eutectic microstructure agreeing well with the experimental and theoretical study [13] and act as a robust basis for further investigation of the influence of interface instability on the pattern continuity.

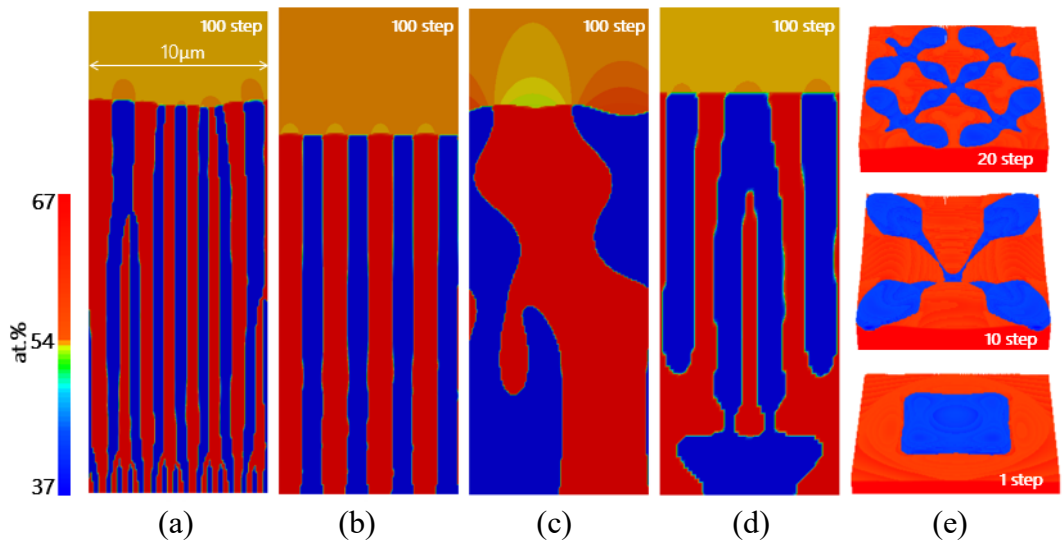

Fig.4 Simulation results generated by concentration field showing spacing transition for different value of initial lamellar spacing $\lambda_{ini}$= (a) 0.625, (b) 2.5, (c)(d)(e) 10 μm.  (a) After some rods are overgrown in the beginning, the steady-state spacing becomes larger. (b) All lamellae can survive and spacing keep unchanged. (c) Lamella grow in an inclined angle and spacing keeps fluctuating. (d)(e) In the 3D model extended from (c), a intial single rod splits into several rods.

## 4. Results

*4.1 Microstructure simulation in three-dimension (3D)*

The simulation of microstructure in three dimensions (3D) has been conducted. Since the initial nuclei in the simulation box are dense and have very small mutual distance, the lamellar spacing in each velocity increases as the solid-liquid interface moved forward. In Fig.5, the 3D simulation boxes at steady state (first row), their cross-section images of growth direction (second row) and the corresponding experimental results (third row) are compared. In the simulation results, it can be discerned that higher interface velocity gives rise to finer microstructure in accordance with the experimental results. Fig.6 compares the lamellar spacing in experiments, JH solution, and PF simulation. All results show the similar decreasing tendency of lamellar spacing as interface velocity increases, but the difference between them is still noticeable. As summarized in Table 3, under three interface velocities (10 mm·h$^{-1}$, 50 mm·h$^{-1}$, 100 mm·h$^{-1}$), the values of lamellar spacing in PF simulation deviate from the experimentally measured ones in the range of 21% to 30%. This suggests that the physical parameters we have chosen still have shortages and can be further optimized. Moreover, a significant number of preceding studies have confirmed that realistic eutectics grow within a range near the extreme condition, rather than precisely at the extreme condition with lowest undercooling [19][26]. Thus, by choosing more appropriate parameters, the deviations can be diminished but may not be eliminated completely. On the other hand, Fig.5 shows the pattern in simulation results is visually similar to that of experiment images having flat solid-solid interfaces and inclined rods. The reproduced similarity was realized by incorporation of a four-fold anisotropy and fixed initial orientation angle for each $Mo_5Si_3$ nucleus, not by a rigorous description of solid-solid interface energy claimed to have an anisotropy caused by lattice mismatch. This compromise was made due to the limitation in manipulating the anisotropy energy function between solid phases in this model. The effect of anisotropy of solid-solid interface energy will be further discussed in another phase-field model afterward.

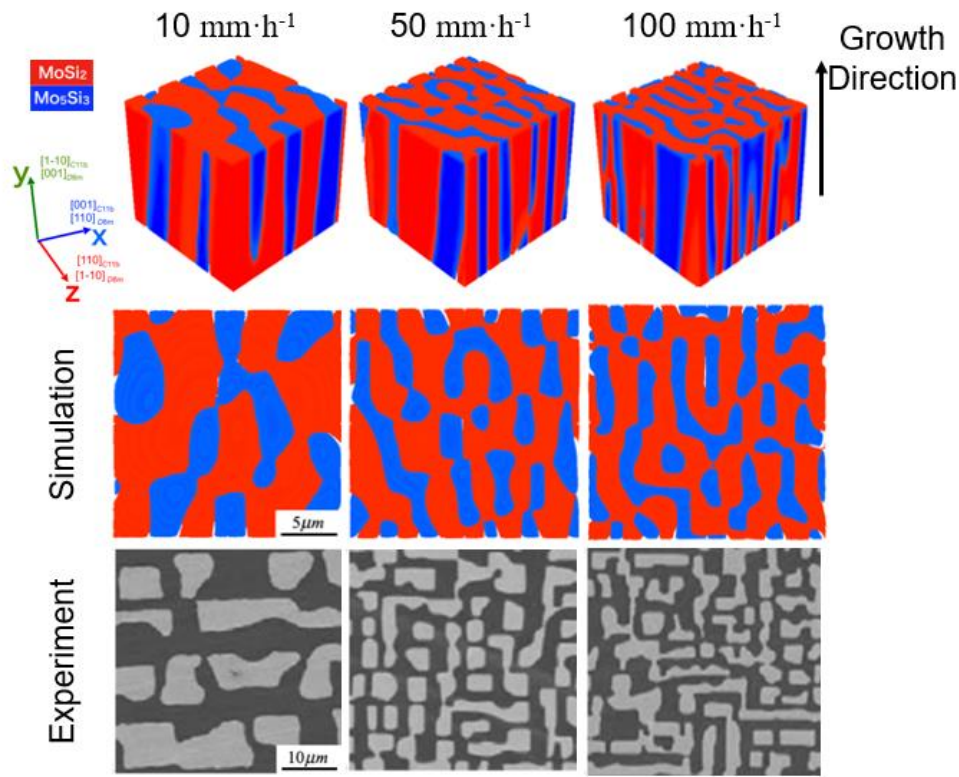

Fig.5 Comparison of microstructure between phase-field simulation (upper row: 3D bird'-eye view, center row: horizontal cross-section, red:$MoSi_2$, blue:$Mo_5Si_3$) results and experimental results (bottom row, dark:$MoSi_2$, grey:$Mo_5Si_3$ after Matsunoshita et al. [6])) for three growth rates of 10 mm·h$^{-1}$(left), 50 mm·h$^{-1}$ (center) and 100 mm·h$^{-1}$(right).

in the simulation box are dense and have very small mutual distance, the lamellar spacing in each velocity increases as the solid-liquid interface moved forward. In Fig.5, the 3D simulation boxes at steady state (first row), their cross-section images of growth direction (second row) and the corresponding experimental results (third row) are compared. In the simulation results, it can be discerned that higher interface velocity gives rise to finer microstructure in accordance with the experimental results. Fig.6 compares the lamellar spacing in experiments, JH solution, and PF simulation. All results show the similar decreasing tendency of lamellar spacing as interface velocity increases, but the difference between them is still noticeable. As summarized in Table 3, under three interface velocities (10 mm·h$^{-1}$, 50 mm·h$^{-1}$, 100 mm·h$^{-1}$), the values of lamellar spacing in PF simulation deviate from the experimentally measured ones in the range of 21% to 30%. This suggests that the physical parameters we have chosen still have shortages and can be further optimized. Moreover, a significant number of preceding studies have confirmed that realistic eutectics grow within a range near the extreme condition, rather than precisely at the extreme condition with lowest undercooling [19][26]. Thus, by choosing more appropriate parameters, the deviations can be diminished but may not be eliminated completely. On the other hand, Fig.5 shows the pattern in simulation results is visually similar to that of experiment images having flat solid-solid interfaces and inclined rods. The reproduced similarity was realized by incorporation of a four-fold anisotropy and fixed initial orientation angle for each $Mo_5Si_3$ nucleus, not by a rigorous description of solid-solid interface energy claimed to have an anisotropy caused by lattice mismatch. This compromise was made due to the limitation in manipulating the anisotropy energy function between solid phases in this model. The effect of anisotropy of solid-solid interface energy will be further discussed in another phase-field model afterward.

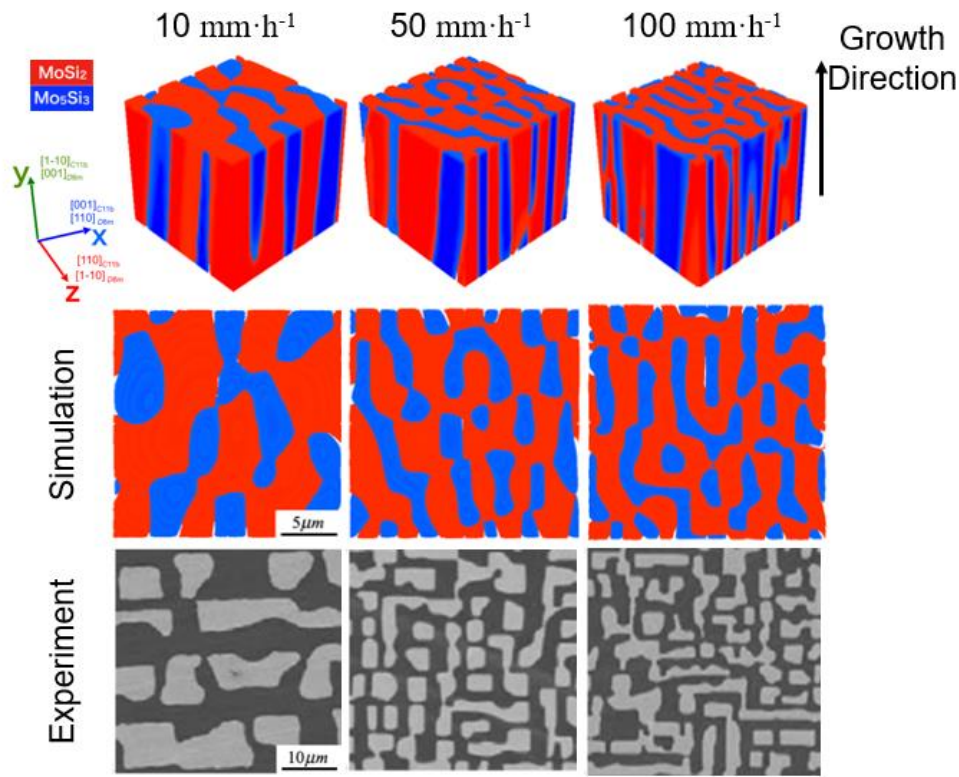

Fig.5 Comparison of microstructure between phase-field simulation (upper row: 3D bird'-eye view, center row: horizontal cross-section, red:$MoSi_2$, blue:$Mo_5Si_3$) results and experimental results (bottom row, dark:$MoSi_2$, grey:$Mo_5Si_3$ after Matsunoshita et al. [6])) for three growth rates of 10 mm·h$^{-1}$(left), 50 mm·h$^{-1}$ (center) and 100 mm·h$^{-1}$(right).

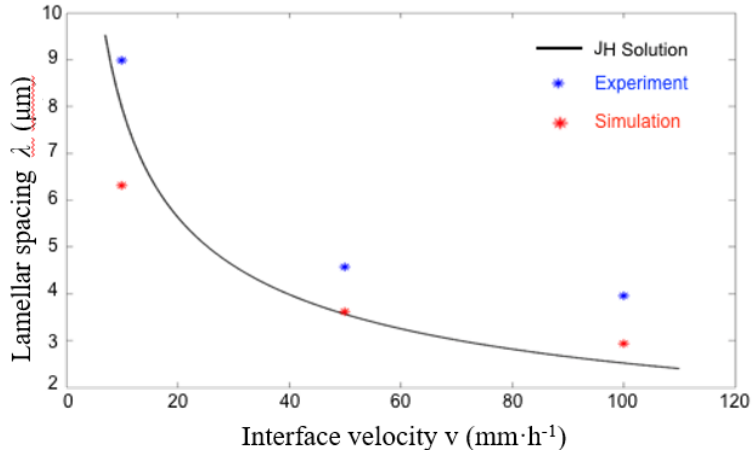

Fig.6 Comparison of lamellar spacing evaluated from Jackson-Hunt solution, experiment and phase field simulation. All of them show the same decreasing tendency as the interface velocity increases.

Table 3 Comparison of lamellar spacing results from experiment and three-dimension phase-field (PF) simulation.

|  | Interface Velocity, $v_{int}$ (mm·h$^{-1}$) | | |
|---|---|---|---|
|  | 10 | 50 | 100 |
|  | Lamellar Spacing, μm | | |
| Experiment | 8.98 | 4.58 | 3.95 |
| PF Simulation | 6.31 | 3.61 | 2.93 |
| Deviation | 29.8% | 21.1% | 25.7% |

*4.2 Interface instability and lamellar discontinuity*

As mentioned in the Introduction, the solid-liquid interface instability during solidification contributes to the discontinuous lamellar pattern. Basically, this instability involves the emergence of new growth sites of the Mo$_5$Si$_3$ rod phase. The manner of generating new growth sites can be either branching (Fig.4d) or nucleation ([27], Fig.8). Owing to the accumulation of rejected solute atoms, the concaves of a lamella interface are formed. Subsequently, the undercooling in the concave for the counter phase becomes larger, making it possible for the nucleation. Despite the fact that there is still no evidence to exclude the branching mechanism of Mo$_5$Si$_3$ phase, in this study, nucleation is considered as a representation of the interface instability because of the inspiration from [27]. The incorporation of continuous nucleation over the solid-liquid interface is expected to act as a footstep to understand how the interface instability affects the pattern continuity.

 Three groups of simulation have been performed to investigate the effect of nucleation. In Case 1, it is assumed that no nucleation occurred. Fig.7 shows the 3D simulation domain, horizontal cross sections, the graph of evolving lamellar spacing, and the longitudinal cross-section. It can be observed that right from the start, the spacing remains to be relatively stable. Later on, the spacing goes up as some rods are overgrown by the matrix. Eventually, the spacing reaches a larger, relatively stable value. Next, in Case 2, as shown in Fig.8, nucleation is introduced. For every 0.05 second, one nucleus is put at a place over the solid-liquid interface where there has a largest undercooling. It should be noted that although the temperature is uniform over the interface, the concentration is inhomogeneous and contributes to the varying undercooling over the interface. As a result, the discontinuity is reproduced seeing from the longitudinal cross-section image (Fig.8d), and the insight of the formation mechanism can be obtained from the graph of spacing-time in Fig.8c. After an initial ascending, the lamellar spacing keeps fluctuating over a long period of time. This sinusoidal-like variation of spacing is the consequence of alternate dominance of nucleation and overgrowth: right after a new nucleus appears over the interface, the spacing decreases because of the increased number of rods. Later on, the new nucleus grows further into a rod, then replaces or merges with the adjacent rods; subsequently, the number of rods decreases and the spacing increases. The repetition of this process gives rise to the fluctuation of lamellar spacing and the discontinuity of the pattern. Finally, in Case 3, as shown in Fig.9, the frequency of nuclei addition is increased to every 0.02 second. The resulting pattern becomes more discontinuous and even intersected. It is interesting that a similar pattern has been observed in the $MoSi_2/Mo_5Si_3$ eutectic with a ternary composition of Mo-54at%Si-0.1at% Co [7].

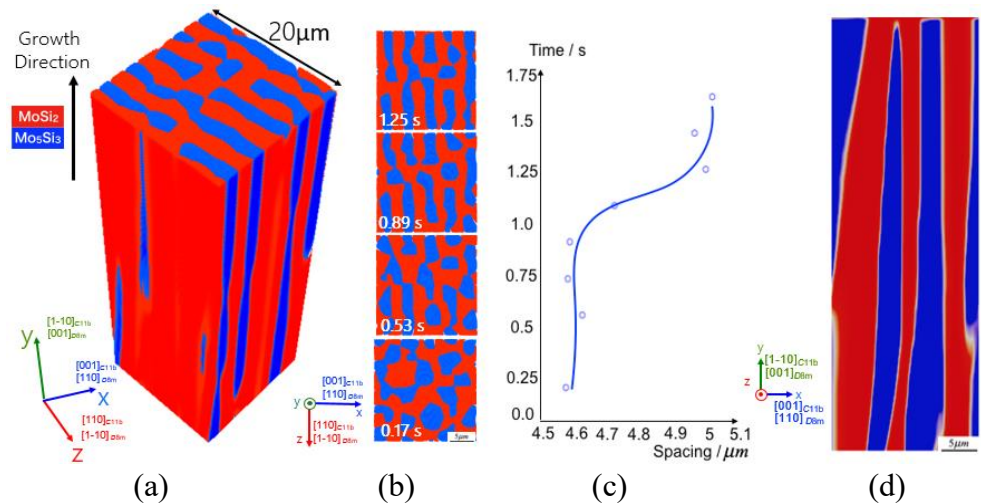

(a) (b) (c) (d)

Fig.7 Results of three-dimension simulation without nucleation. (a) $Mo_5Si_3$ rod (blue) or lamellae in the $MoSi_2$ matrix (red); (b) the microstructure of cross section is evolving along with time; (c) lamellar spacing keeps relatively stable at the beginning, then becomes larger and reaches a steady state; (d) The overgrowth of rod contributes to the increase of spacing.

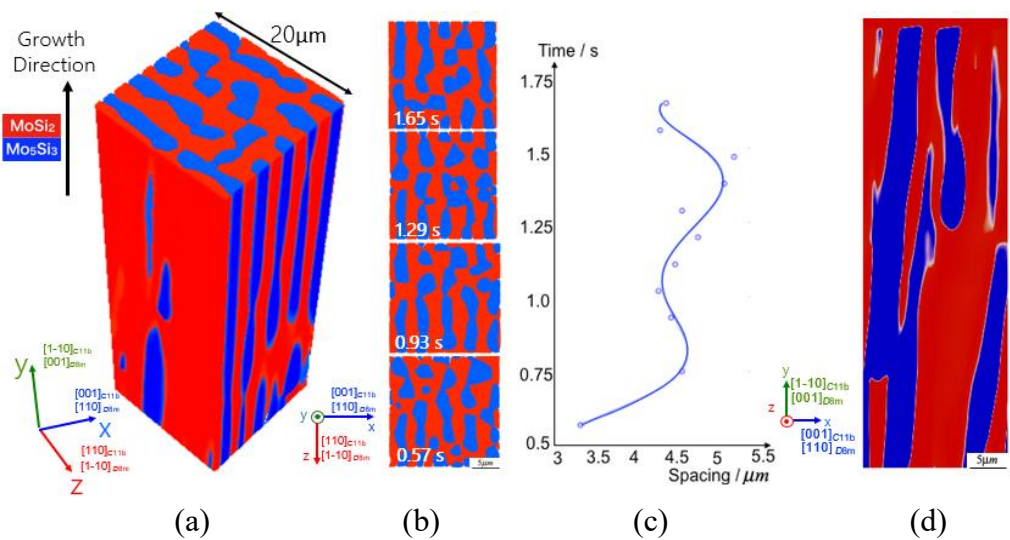

(a) (b) (c) (d)

Fig.8 Results of nucleation-introduced three-dimension simulation. The frequency of nucleation is at 20 s$^{-1}$. (a) $Mo_5Si_3$ rod (blue) or lamellae in the $MoSi_2$ matrix (red); (b) the microstructure of cross section is evolving along with time; (c) after a short ascending in the beginning, the lamellar spacing

C. Zhu et al. (2019)keep fluctuating over a long period; (d) emergence of new grain (nucleation) and elimination of old grain contribute to the formation of discontinuity.

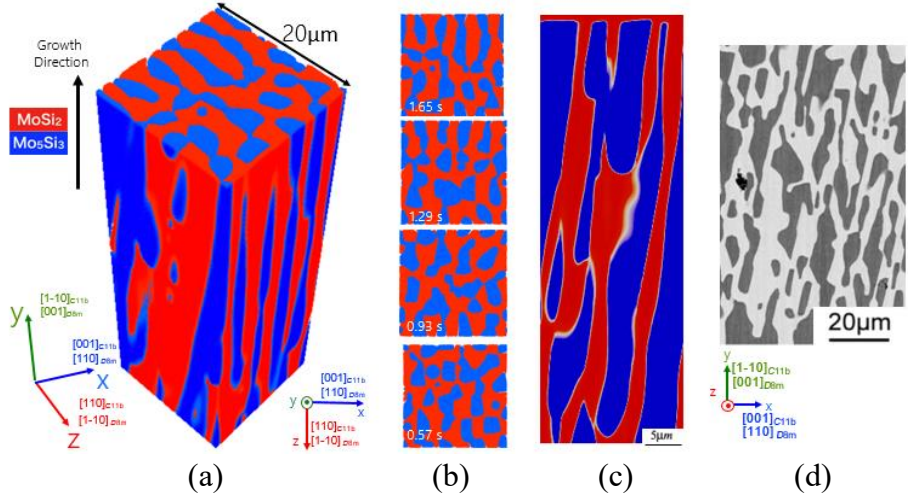

(a) (b) (c) (d)

Fig.9 Results of nucleation-introduced three-dimension simulation. The frequency of nucleation is increased to 50 s$^{-1}$. (a-c) the simulated pattern became more discontinuous and intersected; (d) Experimentally observed pattern in MoSi$_2$/Mo$_5$Si$_3$ eutectic alloyed with 0.1 at% Co (after Fujiwara et al. [7]).

*4.3 Lattice misfit and interface energy anisotropy*

In Fig.1 (after Fujiwara et al. [7]), we can see the Mo$_5$Si$_3$ rod in the MoSi$_2$ has a cuboidal shape with four flat interfaces, a pair of which is inclined from the growth direction. As expected by Wulff's construction [30][31], the equilibrium shape should have the lowest surface energy. Thus, knowing the shape of the Mo$_5$Si$_3$ rod, it is straightforward to construct a corresponding anisotropic function of MoSi$_2$/Mo$_5$Si$_3$ interface energy. As schematically illustrated in Fig.10a, a similar cuboidal equilibrium shape is enveloped in an anisotropy function with four cusps. It should be noted that the anisotropic function also has an inclination, corresponding to the interface inclination. In Fig.10b, a 2D plot (Eq.9) with inclined cusps represents the anisotropy of interface energy on the vertical cross section. In order to get a complete description of the interface energy anisotropy, the remaining questions concerning the specifications of the 2D anisotropy function are the following two: "What is the inclination angle?", "How deep should the cusp be?"

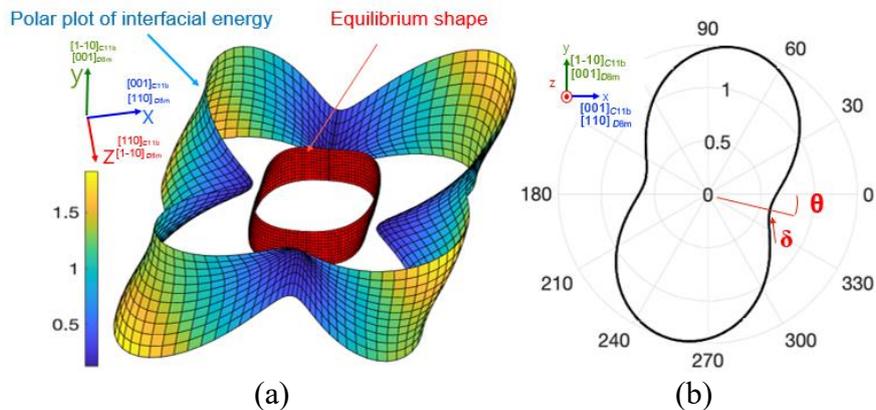

(a) (b)

Fig.10 Anisotropic description of the MoSi$_2$/Mo$_5$Si$_3$ interface energy based on the characteristics of the script lamellar pattern. (a) 3D surface of anisotropic function with cusps and its equilibrium shape

1313

inside; (b) plot of a two-dimension anisotropic function which is a reduced and simplified version of the 3D surface. It also has cusps which can give rise to the inclined interfaces. $\theta$ is the inclination angle and $\delta$ is the anisotropic strength.

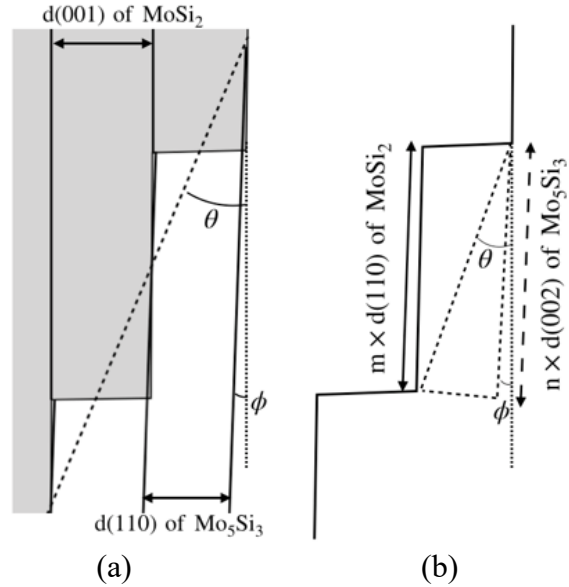

(a)          (b)

Fig.11 Schematics of lattice misfit accommodation by the formation of ledge-terrace structure. (a) elimination of the ledge misfit between (001) of $MoSi_2$ and (110) of $Mo_5Si_3$ by a rotation of angle $\phi$; (b) elimination of terrace misfit between (110) plane of $MoSi_2$ and (002) plane of $Mo_5Si_3$ by insertion of dislocations. The resulting inclination of interface is denoted by angle $\theta$. See Ref. [7] for the details of this crystallographic analysis.

To answer the first question, crystallographic analysis and experimental observation should be considered. Based on lattice parameters at the eutectic temperature (1900°C), as illustrated in Fig.11, the mismatch between the two crystals of $MoSi_2$ and $Mo_5Si_3$ can be eliminated by formation of ledge-terrace structure with a set of the atomic arrangement parameters, including the inclination angle $\theta$ having a value of 13.9° [7]. In the experimental observation by high-resolution TEM (Fig.1b), there is also a ledge-terrace structure at the interface, which has an inclination angle of 13.5°, almost equal to the analysis result. This consistency between analysis and experiment confirms that the inclined $MoSi_2/Mo_5Si_3$ interface has the lowest lattice strain energy and corresponds to the cusp in the anisotropy function constructed in Fig.10b. Therefore, the inclination angle of the function is determined to be 13.5°, the same value as experimental measurement. Also, it is worth noting that the interface with anisotropic energy we are referring to is the macroscopic interface on the micrometer scale rather than the ledge or terrace interface on the nanometer scale. In the present study, the interface energy includes both components of elastic energy due to the coherency (i.e., due to crystallographic structure) and chemical energy due to interatomic bonding (i.e., electronic structure).

Having already constructed a cusp anisotropy function with a 13.5° inclination angle, we want to know if this anisotropy representation of the interface energy can enable the interface to grow precisely at the 13.5°angle. Our tool is the phase-field model constructed from Eqs.8, 9 with anisotropic interface energy specified by the anisotropy functions (Eqs.10, 11). The initial condition of the microstructure has three phases with a triple junction and three interfaces. As shown in Fig.12a, the solid/solid interface has anisotropic interfacial energy (AIE) expressed by Eq.10, and the other two (i.e.,

liquid/MoSi$_2$ interface and liquid/Mo$_5$Si$_3$ interface) have isotropic interface energy (IIE). Since the model is based on the principle of energy minimization, it is expected that the AIE will results in the formation of MoSi$_2$/Mo$_5$Si$_3$ interface along the orientation perpendicular to the cusps. However, there is a force equilibrium at the triple junction, and therefore, the inclination angle of the interface deviates from the angle expected from the AIE. This indicates that the cusp needs to be deeper than a critical value, i.e., the AIE needs to be strong enough, to be able to form an interface with inclination angle expected from the cusp of AIE.

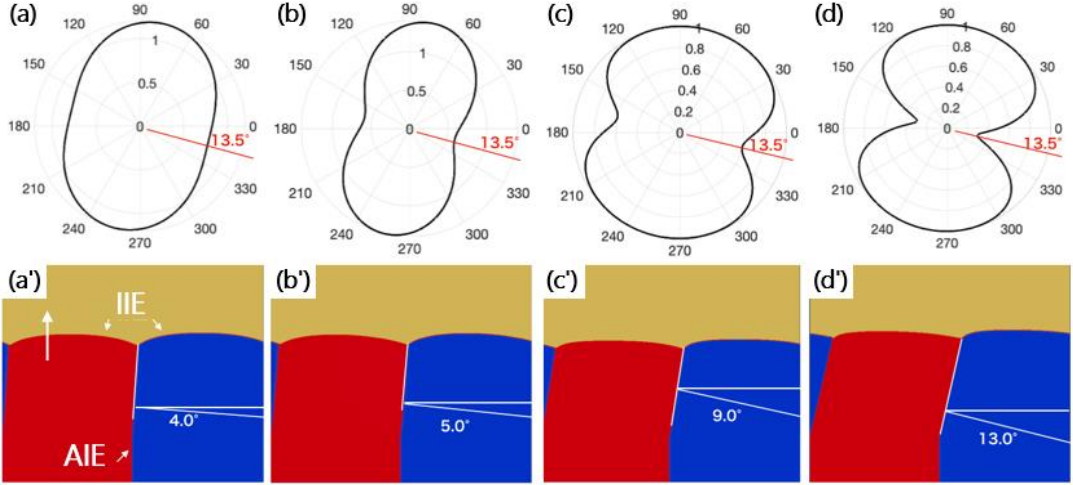

Fig.12 (a-d) Four types of two-fold anisotropy functions used to investigate the effect of cusp depth on the inclination angle of the solid-solid interface, and (a'-d') corresponding simulation results: (a,a') δ=0.2 in Eq.10, (b, b') δ=0.4 in Eq.10, (c,c') δ=0.4, w=0.4 in Eq.11, (d,d') δ=0.7, w=0.4 in Eq.11.

As shown in Fig.12a, the cusps are shallow or nearly flat, the interface only inclines in a small angle of 4.0°, being much smaller than 13.5°. After increasing the anisotropic strength and deepening the cusp (Fig.12b), the resulting inclination becomes larger and increases to 5°, still being far below 13.5°. These results confirm that the liquid-solid interface does affect the inclination angle, and increasing the anisotropy strength will make the inclination angle closer to the expected value. Next, we use another anisotropy function (Eq.11) with sharper cusps. In Fig.12c, the resulting angle of the interface is 9.0°, being relatively close to 15°. After further increasing the strength in Fig.12d, the angle increases to be 13.0°, nearly at the value of 13.5°. However, when the anisotropy strength is further increased, numerical instability occurs, and no convergence of the inclination angle can be reached.

## 5. Discussions

It can be claimed that the origin of the interface instability is due to the complex crystal structure for both MoSi$_2$ and Mo$_5$Si$_3$ phases. In a crystallographic analysis [7], MoSi$_2$ and Mo$_5$Si$_3$ have body-centered tetragonal structures with the types of $C11_b$ and $D8_m$, respectively. The $(1\bar{1}0)$ plane of MoSi$_2$ and (0 0 1) plane of Mo$_5$Si$_3$ are found to be nearly perpendicular to the temperature gradient or growth direction. The formation of this set of crystallographic orientations is basically attributed to the directional bonding of intermetallic compounds. These compounds have a high degree of order after the reorganization from the liquid state (i.e., solidification), and atoms can only attach to the interface to form bonds with the crystal along well-defined crystallographic orientation, resulting in a high anisotropy of interface mobility. Accordingly, the two phases cannot grow in a coupled manner even with a slight difference of the interface mobilities along the growth directions. Therefore, a persistent interface instability occurs during the solidification process, acting as the basis of our assumption of continuous nucleation ahead of the solid-liquid interface. In addition, the nucleation process may be influenced by the alloying element. Similar to the result in Fig.9, it has been reported that very small amount Cr (typically 0.1mass%) to Al-Zn significantly reduces the grain size by changing the nucleation condition. The mechanism involves the transformation from icosahedral short-range order (ISRO) in the liquid to quasicrystals, which implies smaller interface energy and lower nucleation barrier [28][29].

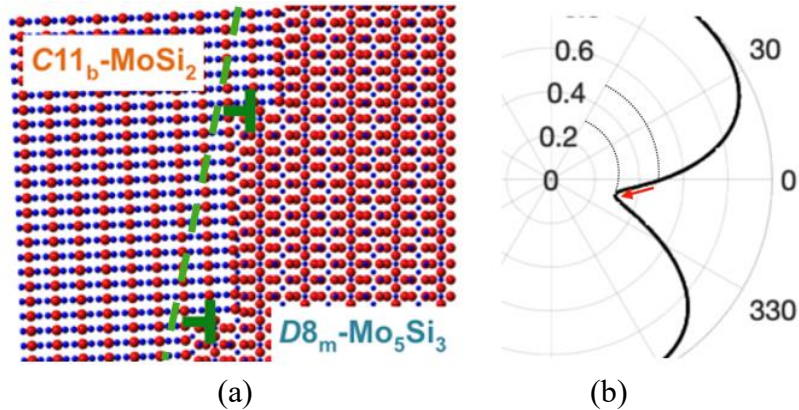

(a)                                    (b)

Fig.13 (a) Schematic illustration of the atomic arrangement of ledge-terrace structure; (b) the energy slump due to the inclination of macroscopic interface and formation of ledge-terrace structure.

The crystallographic analysis demonstrates that the significant lattice misfit can be eliminated by forming a ledge-terrace structure, which has also been observed in the experiment. The results of inclination angle and atomic arrangement in analysis and experiment are almost consistent, which means the effect of solute diffusion in the liquid phase and force equilibrium at the triple junction play the immaterial roles. Thus, it becomes clear that misfit accommodation plays a dominated role in determining morphology. Furthermore, the phase-field simulation result shows that the inclination angle can be nearly locked at the predetermined value only when the cusp is very sharp and deep. This means that energy relaxation accompanying with the lattice accommodation is strong, even to the extent of an energy slump (i.e., fast decrease). As illustrated in Fig.13, by the formation of inclined morphology rather than a vertical one parallel to the temperature gradient, the rotation of 13.5° makes the state of interface energy decrease from the level of 0.5 to a lower level 0.3. The formation of this slump has a close relationship with the coherency state of the interface and involves the generation of the elastic strain energy in the interface region. Since the current model uses a macroscopic interface description, the future work could be modeling on a microscopic scale with a rigorous consideration of elastic energy in the phase-field model.

## 6. Conclusions

In this study, we have discussed the formation mechanism of the script lamellar structure found in the directionally-solidified $MoSi_2/Mo_5Si_3$ eutectic structure. The conclusions can be summarized as follows:

(I)   The phase-field model is able to reproduce the essential physics during eutectic solidification, including interface undercooling, overgrowth, splitting, and dimension-restricted (diffusion-

(induced) inclination.

(II)  The 3D eutectic simulation reproduces the process of spacing selection and shows a length scale of microstructure close to that of experiment results. This demonstrates the validity of the present phase field model and the set of physical parameters.

(III) Instability of the solid-liquid interface is responsible for the discontinuity of script lamellar pattern. A mechanism of continuous nucleation ahead of the solid-liquid interface has been proposed. The alloying element is claimed to be able to affect the nucleation condition, accordingly, the discontinuity of the irregular eutectic pattern.

(IV) The interface inclination is caused by crystallographic interaction between the solid phases and has almost nothing to do with other factors such as solute diffusion in the liquid phase or the surface tension of the solid-liquid interface. The formation of ledge-terrace structure leads to a substantial relaxation of the misfit strain energy, which can be characterized as a slump of energy generated in the region of $MoSi_2/Mo_5Si_3$ boundary.

**Acknowledgment**

This study was supported by the Advanced Low Carbon Technology Research and Development Program of the Japan Science and Technology Agency. This study was partly supported by the Center for Computational Materials Science, Institute for Materials Research, Tohoku University and the Cyber Science Center, Tohoku University.

**Appendix Table** Parameters used in the phase-field model with anisotropic interface energy

| | **Physical Parameters** | | |
|---|---|---|---|
| Gas constant | $R$ | 8.314 | $J \cdot K^{-1} \cdot mol^{-1}$ |
| Grid size | $L_0$ | $4 \times 10^{-8}$ | m |
| Temperature | $T_0$ | 1000 | K |
| Molar volume | $V_0$ | $7 \times 10^{-6}$ | $m^3 \cdot mol^{-1}$ |
| Interface energy | $\gamma$ | 0.5 | $J \cdot m^{-2}$ |
| Interface width | $\delta$ | $7.0 \times L_0$ | m |
| Mobility | $m$ | 1.0 | [-] |
| | **Parameters in the phase-field model** | | |
| Gradient coefficient | $\varepsilon_{ij} = \begin{bmatrix} 0 & \varepsilon(\theta) & \varepsilon_0 \\ \varepsilon(\theta) & 0 & \varepsilon_0 \\ \varepsilon_0 & \varepsilon_0 & 0 \end{bmatrix}$ | $\bar{\gamma} = \dfrac{\gamma V_0 R}{T_0 L_0}, \varepsilon_0 = \dfrac{8\delta\bar{\gamma}}{\pi^2}$ $\varepsilon(\theta) = \varepsilon_0 \cdot a(\theta)$ | |
| Double well penalty | $W_{ij} = \begin{bmatrix} 0 & W_0 & W_0 \\ W_0 & 0 & W_0 \\ W_0 & W_0 & 0 \end{bmatrix}$ | $W_0 = \dfrac{4\bar{\gamma}}{\delta}$ | |
| Phase field mobility | $M_{ij} = \begin{bmatrix} 0 & 0 & M_0 \\ M_0 & 0 & M_0 \\ M_0 & M_0 & 0 \end{bmatrix}$ | $M_0 = \dfrac{m\pi^2}{8\delta}$ | |
| Driving force | $\Delta f_{ij} = \begin{bmatrix} 0 & 0 & \Delta f_0 \\ & 0 & \Delta f_0 \\ -\Delta f_0 & -\Delta f_0 & 0 \end{bmatrix}$ | $\Delta f_0 = \dfrac{15.0 \ J \cdot mol^{-1}}{RT}$ | |
| Gird size | $\Delta x$ | 1.0 | |
| Time step | $\Delta t$ | 5.0 | |
| Domain size | $N_x \times N_y$ | 200×200 | |